\documentclass[epj]{svjour}
\usepackage{latexsym}
\usepackage{graphicx}

\begin{document}

\title{Low temperature magnetic phase diagram of the cubic non-Fermi liquid system CeIn$_{3-x}$Sn$_x$}
\author{P. Pedrazzini\inst{1} \and M. G\'omez Berisso\inst{1} \and N. Caroca-Canales\inst{2} \and M. Deppe\inst{2} \and C. Geibel\inst{2} \and J. G. Sereni\inst{1}
}                     
\offprints{Pablo Pedrazzini, e-mail: pedrazp@cab.cnea.gov.ar}   
\institute{Lab.\ Bajas Temperaturas, Centro At\'omico Bariloche (CNEA), 8400 S.C. de Bariloche, Argentina \and Max-Planck Institute for Chemical Physics of Solids, D-01187 Dresden, Germany}
\date{Received: date / Revised version: date}
%
\abstract{
In this paper we report a comprehensive study of the magnetic susceptibility ($\chi$), resistivity ($\rho$), and specific heat ($C_{\rm P}$), down to 0.5\,K of the cubic CeIn$_{3-x}$Sn$_x$ alloy. The ground state of this system evolves from antiferromagnetic (AF) in CeIn$_3$ ($T_{\rm N}=10.2\,$K) to intermediate-valent in CeSn$_3$, and represents the first example of a Ce-lattice cubic non-Fermi liquid (NFL) system where $T_{\rm N}(x)$ can be traced down to $T=0$ over more than a decade of temperature. Our results indicate that the disappearance of the AF state occurs near $x_{\rm c}\approx 0.7$, although already at $x\approx 0.4$ significant modifications of the magnetic ground state are observed. Between these concentrations, clear NFL signatures are observed, such as $\rho(T)\approx\rho_0+A\,T^n$ (with $n<1.5$) and $C_{\rm P}(T)\propto-T\ln(T)$ dependencies. Within the ordered phase a first order phase transition occurs for $0.25 < x < 0.5$. With larger Sn doping, different weak $\rho(T)$ dependencies are observed at low temperatures between $x=1$ and $x=3$ while $C_{\rm P}/T$ shows only a weak temperature dependence.
} 
\maketitle

\section{Introduction}
\label{intro}

The classical description of heavy fermion (HF) metals in terms of a ``Fermi liquid'' of quasiparticles with a large renormalized mass fails in several Ce, Yb and U systems. During the last decade, many examples of ``non-Fermi liquid'' (NFL) behaviour have been identified by the anomalous dependences of the low temperature thermodynamic and transport properties. Different scenarios have been proposed for these NFL effects, in particular the case of NFL associated to a quantum critical point (QCP) is nowadays the subject of intensive research, both experimental \cite{Stewart01,Lohneysen99,Steglich00} and theoretical \cite{Si01,Coleman02}.

A quantum critical point is a singularity in the phase diagram where a second order phase transition is driven to zero temperature by means of a non-thermal control parameter such as pressure, magnetic field or alloying. In HF systems, the QCP results from the competition of the Kondo effect (the screening of local magnetic moments by conduction electrons) and the RKKY interaction between neighboring magnetic moments.\cite{Continentino00} The nature of the QCP is a
subject of current discussion, but in any case a strong influence of the dimensionality of the spin-fluctuations is expected.\cite{Si01} This raises the question of the influence that structural symmetry may have on NFL behaviour.

At present, most of the systems that have been thoroughly investigated in connection with a QCP have tetragonal or lower structural symmetry, and thus, an intrinsic structural anisotropy. In order to avoid this constraint a cubic system should be studied. To our knowledge, no {\em cubic doped system} has been studied yet in this context. Because of its FCC structure, CeIn$_{3-x}$Sn$_x$, with a cubic symmetry at the Ce site, can be considered an excellent candidate. Previous research has shown that the antiferromagnetic (AF) order present in CeIn$_3$ at $T_{\rm N}=10.2\,$K drops to zero with Sn doping (increasing $x$), and $T_{\rm N}$ was predicted to disappear near $x_{\rm c}\approx 0.4$.\cite{Lawrence79} Furthermore, resistivity measurements performed on stoichiometric CeIn$_3$ under pressure reveal NFL behaviour with a non-quadratic temperature dependence of the resistivity ($\Delta\rho\propto T^{1.6}$) close to the critical pressure $p_{\rm c}\approx 26$\,kbar where the ordering temperature extrapolates to zero.\cite{Mathur98,Knebel01}

In two short conference contributions \cite{Pedrazzini02,Pedrazzini03} we have presented low temperature specific heat and resistivity results of our research on CeIn$_{3-x}$Sn$_x$ for $x\le 1$. Those preliminary results relocate the critical concentration at $x_{\rm c}\approx 0.65$ and show evidence of NFL behaviour and a complex magnetic phase diagram. In this work we present more detailed information on structural, magnetic, transport and thermal measurements around the QCP and we extend the previous study to the non-ordered region of the phase diagram.

\section{Experimental details and Results}
\label{sec:experimental}

Eighteen polycrystalline samples covering the full concentration range were prepared by arc melting the proper amounts of pure elements in an Argon atmosphere. The starting materials were Ce of grade 5N and In and Sn of grade 6N. In order to insure homogeneity, the resulting buttons were flipped over and remelted several times. At the end of this process the mass loss was found to be negligible (below 0.2 wt.\%). Samples were annealed at $700^{\rm o}$C for 24 hours and then slowly cooled down to $500^{\rm o}$C within 80 hours. Powder X-ray diffraction confirmed the FCC AuCu$_3$ type structure of all samples and revealed no signs of secondary phases. The magnetic dc-susceptibility ($\chi$) was measured between 2\,K and 300\,K in a commercial SQUID magnetometer under a magnetic field $H=10\,$kOe. Electrical resistivity ($\rho$) was measured with the conventional four-contact technique with a Linear Research LR-700 resistance bridge in the 0.5--300\,K temperature range. At low temperature ($T<6\,$K) a small (extrinsic) superconducting signal was suppressed by an $H=10\,$kOe magnetic field (this is probably due to tiny amounts of In--Sn segregation at the surface and grain boundaries of the samples \cite{Sakurai84Chem}). Specific heat ($C_{\rm P}$) measurements were performed in a He$^3$ quasiadiabatic calorimeter by the usual heat-pulse technique, between 0.45\,K and 20\,K.

The lattice parameters of some of the samples are presented in Fig.~1 together with the crystallographic data previously reported in the literature.\cite{Lawrence79,Maury82,Sakurai84} The lattice parameter ($a$) of the AuCu$_3$-cubic structure increases monotonously with increasing $x$. Between $x=0$ and $x\approx 2$ the data follow the expected Vegard law extracted by comparison between the LaIn$_3$ and LaSn$_3$ lattice parameters. However, a clear deviation from a straight line occurs for $x>2.2$. Such a deviation is associated to a change of the Ce-valence on the Sn rich side due to the reduction of the Ce-$4f$-level occupancy from $n_f=1$ (CeIn$_3$) to $n_f<1$ (CeSn$_3$).\cite{Lawrence79}

\begin{figure}[htb]
\begin{center}
\includegraphics[angle=270,width=.47\textwidth]
{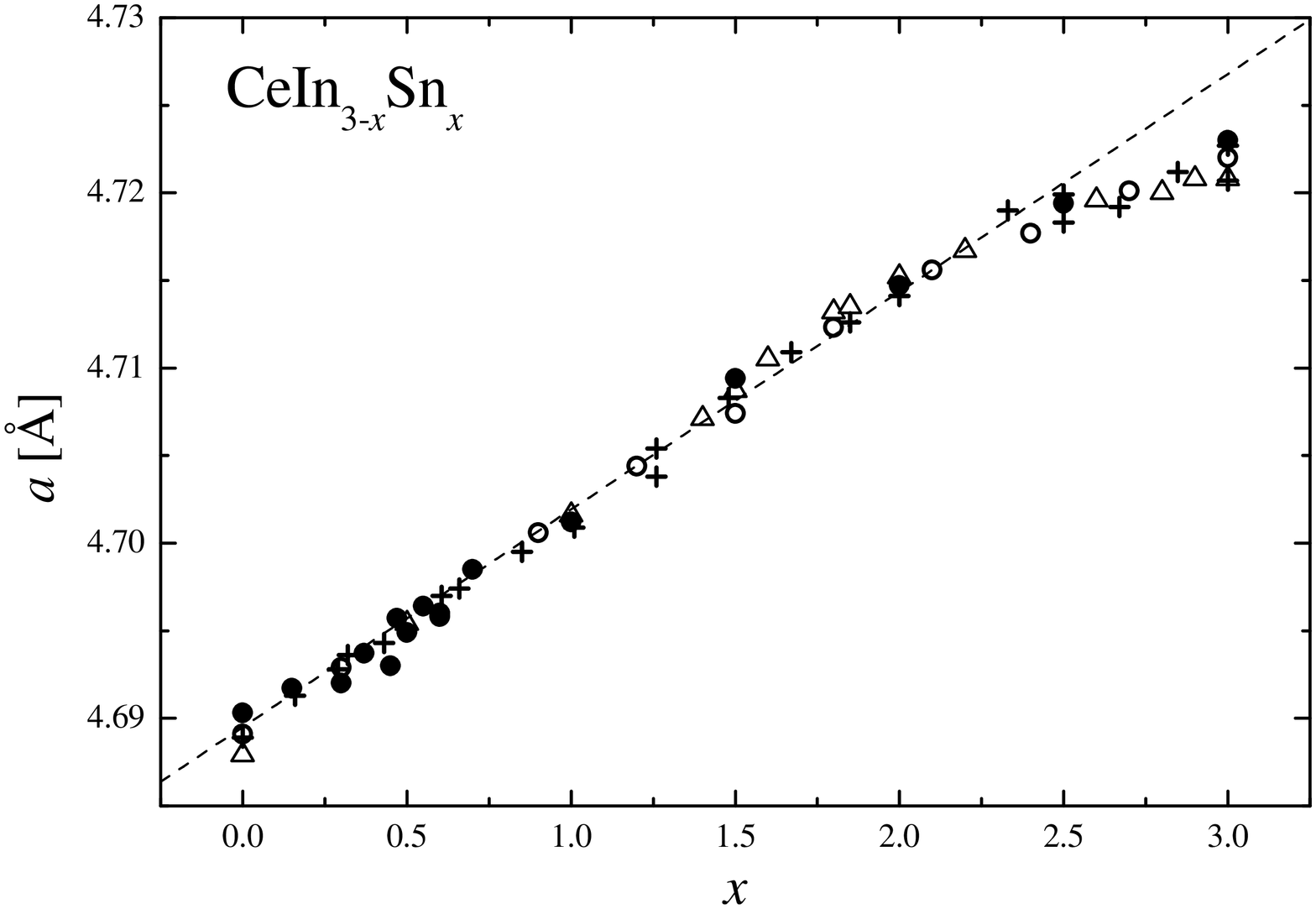}
\caption[]{Lattice parameter ($a$) of the AuCu$_3$-type cubic structure. Our data ($\bullet$) are plotted together with the data taken from references \cite{Lawrence79} ($+$), \cite{Maury82} ($\triangle$) and \cite{Sakurai84} ($\circ$). The line represents Vegard law obtained by the interpolation between LaIn$_3$ and LaSn$_3$ lattice parameters (displaced).}
\end{center}
\label{fig:parametros}
\end{figure}

The magnetic susceptibility is employed to follow the ordering transition (at low $x$) and to study the formation of the intermediate valence (IV) state (for $x>2$). Our $\chi(T)$ data are plotted in Fig.~2 along with the data of pure CeSn$_3$ (taken from Ref.~\cite{Sereni80}). Within our experimental range ($T\geq 2\,$K), the AF transition can be observed for $x<0.45$. At this Sn concentration, $x=0.45$, $\chi_{2\,{\rm K}}(x)\equiv\chi(T=2\,{\rm K},x)$ reaches its maximum value and then it rapidly decreases for $x>0.6$, as shown in the inset of Fig.~2. This decrease is not monotonous because $\chi_{2\,{\rm K}}$ shows a weak increase at $x=2.0$ with respect to the neighbouring samples ($x=1.5$ and 2.5, see the inset of Fig.~2). A weak shoulder around 50\,K is observable at low Sn concentration, which becomes more pronounced for the $x=0.5$ and $0.6$ samples. This deviation from the Curie-Weiss law, already observed below 100\,K, is related to the thermal depopulation of the quartet ($\Gamma_8$) excited level. Eventually, the shoulder evolves into a broad maximum centered at $T^{\chi}_{max}\approx 40\,$K for $x\geq 1$. Coincidentally with the deviation of the unit-cell volume ($V$) from Vegard's law at $x\approx 2$, $T^{\chi}_{max}$ shifts to higher temperature for $x>2$, indicating the increase of the $4f$-conduction state hybridization strength. Once its value exceeds the crystal-field splitting, the system behaves like an IV system, as in pure CeSn$_3$ where $T^{\chi}_{max}\approx 140$\,K. Our $\chi(T)$ results are in good agreement with those reported in the literature for $T>10\,$K.\cite{Dijkman80}

\begin{figure}[htb]
\begin{center}
\includegraphics[angle=270,width=.47\textwidth]
{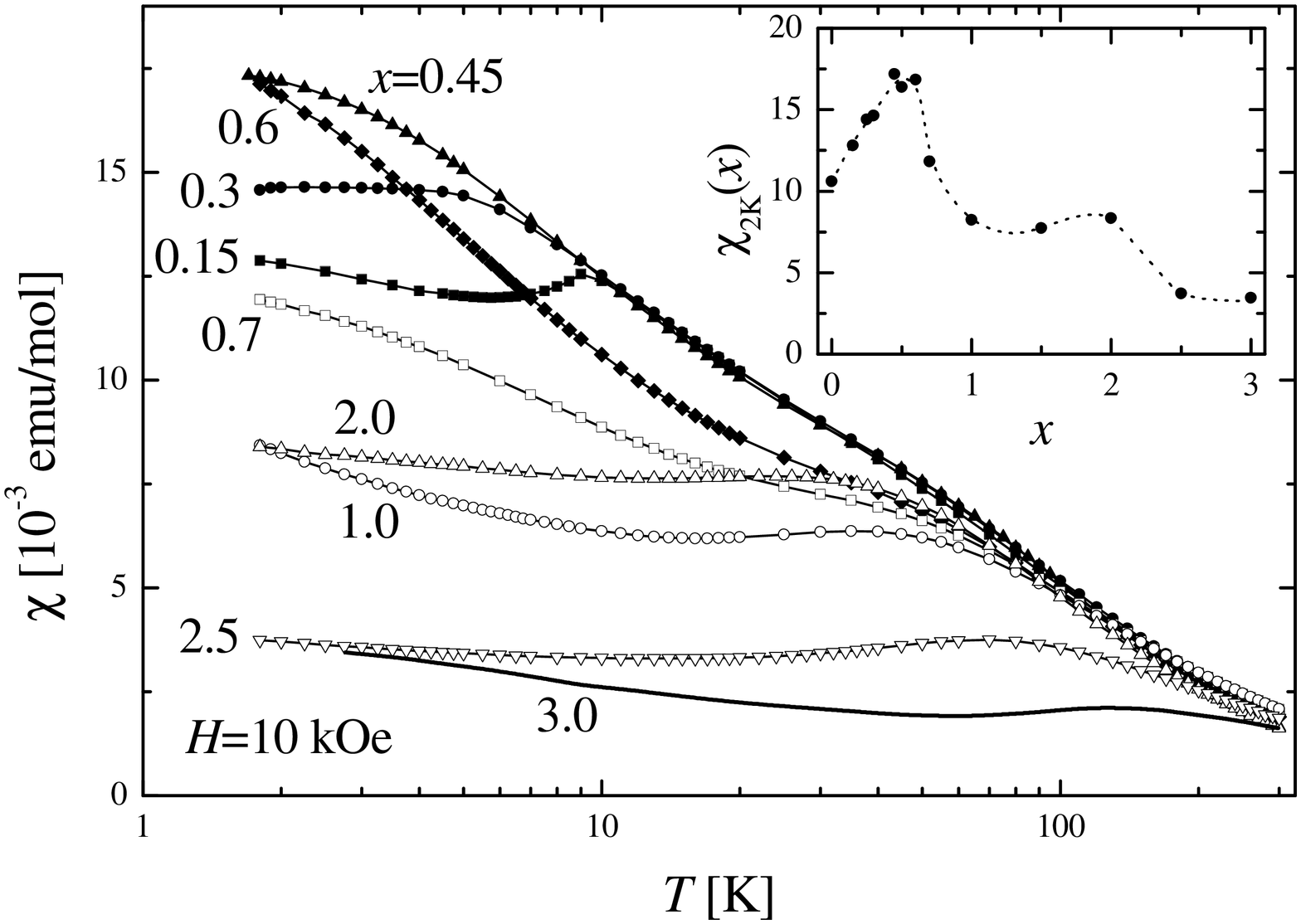}
\caption[]{Magnetic susceptibility of some selected concentrations, plotted against a logarithmic temperature scale. The applied magnetic field was $H=10\,$kOe. Data for CeSn$_3$ is taken from Ref.~\cite{Sereni80}. Inset: $\chi_{2\,{\rm K}}(x)\equiv\chi(T=2\,{\rm K},x)$, the value for $x=3$ is taken from Ref.~\cite{Lawrence79}. The lines are guides to the eye.}
\end{center}
\label{fig:chi}
\end{figure}

Figure 3 displays the electrical resistivity of the studied samples in a logarithmic $T$ scale. For the sake of clarity, the data is normalized as $(\rho(T)-\rho_{170\,{\rm K}})/(\rho_{70\,{\rm K}}-\rho_{170\,{\rm K}})$ for $x\leq 0.7$ in Fig.~3-a, and as $\rho(T)/\rho_{280\,{\rm K}}$ for $x\ge 1$ in Fig.~3-b. The most prominent features extracted from these measurements are: i) the magnetic transition observed in the $0\leq x< 0.6$ range; ii) a maximum in $\rho(T)$ at $T^{\rho}_{max}$ which decreases from 55\,K at $x=0$ down to 20\,K at $x=0.45$, and then increases for $x>0.7$; iii) a minimum at $T\approx 200$\,K resulting from the decrease of the electron-electron scattering and the increase of electron-phonon scattering as the temperature increases.

As shown in Fig.~3-a, the resistivity of the samples with $x\leq 0.7$ merges in a common $\rho(T)$ dependence at high temperature. Furthermore, such a common temperature dependence extends between $T^{\rho}_{max}=20\,$K and nearly room temperature for the samples with $x\geq 0.45$. This fact points to common scattering mechanisms within this range of concentration. At low temperature, the AF transition is characterized by a negative $\rho(T)$ slope in a temperature range around $T_{\rm N}$, followed by the typical decrease in the ordered phase. This effect becomes much weaker at $x=0.25$, and is dominated by a sudden drop due to another transition occuring at $T=T_{\rm I}$ in the $x=0.3 $, $0.37$, $0.41$ and $0.45$ samples. The observation of both a hysteresis in $\rho(T)$ and a peak-like anomaly in $C_{\rm P}(T)$ have led us to conclude that a first order phase transition occurs at $T_{\rm I}$.\cite{Pedrazzini03} This phase transition is no longer detected in the $x=0.5$ sample within our experimental temperature window ($T>0.5\,$K). Instead, a continuous increase in $\rho(T)$ is found down to the lowest temperature, which we again associate to an AF ordering. This behaviour, observed in the resistivity around $T_{\rm N}$ in the $0.15\le x\le 0.5$ concentration range, can be ascribed to the opening of the superzone gap due to the AF order, as reported e.g.~in CeSbNi$_y$.\cite{Kim01} For $x=0.6$ and $0.7$, $\rho(T)$ decreases monotonously with $T$ from their common maximum at 20\,K, with a clear sub-quadratic power law behaviour: $\rho(T)=\rho_0+A T^n$, with $n<1.5$. Resistivity measurements performed down to the milikelvin range confirm an exponent $n\approx 1$ for the sample with $x=0.7$.\cite{Custers03}

At higher Sn concentration the $\rho(T)$ maximum at $T^{\rho}_{max}$ shifts again to a higher temperature, with $T^{\rho}_{max}=30\,$K for $x=1$. No maximum is observed in the samples with $x=2$, 2.5 and 3 (see Fig.~3-b), though it appears at a higher temperature once the phonon contribution is substracted. The evolution of $\rho(T)$ is clearly broken at $x=1.5$, where $\rho(T)$ has a monotonous negative slope in the whole temperature range. This unexpected behaviour for a Ce-lattice system was previously reported in the literature at this concentration,\cite{Sakurai84,Elenbaas80} although in a smaller temperature range that did not include the low temperature saturation value. 

\begin{figure}[htb]
\begin{center}
\includegraphics[angle=0,width=.37\textwidth]
{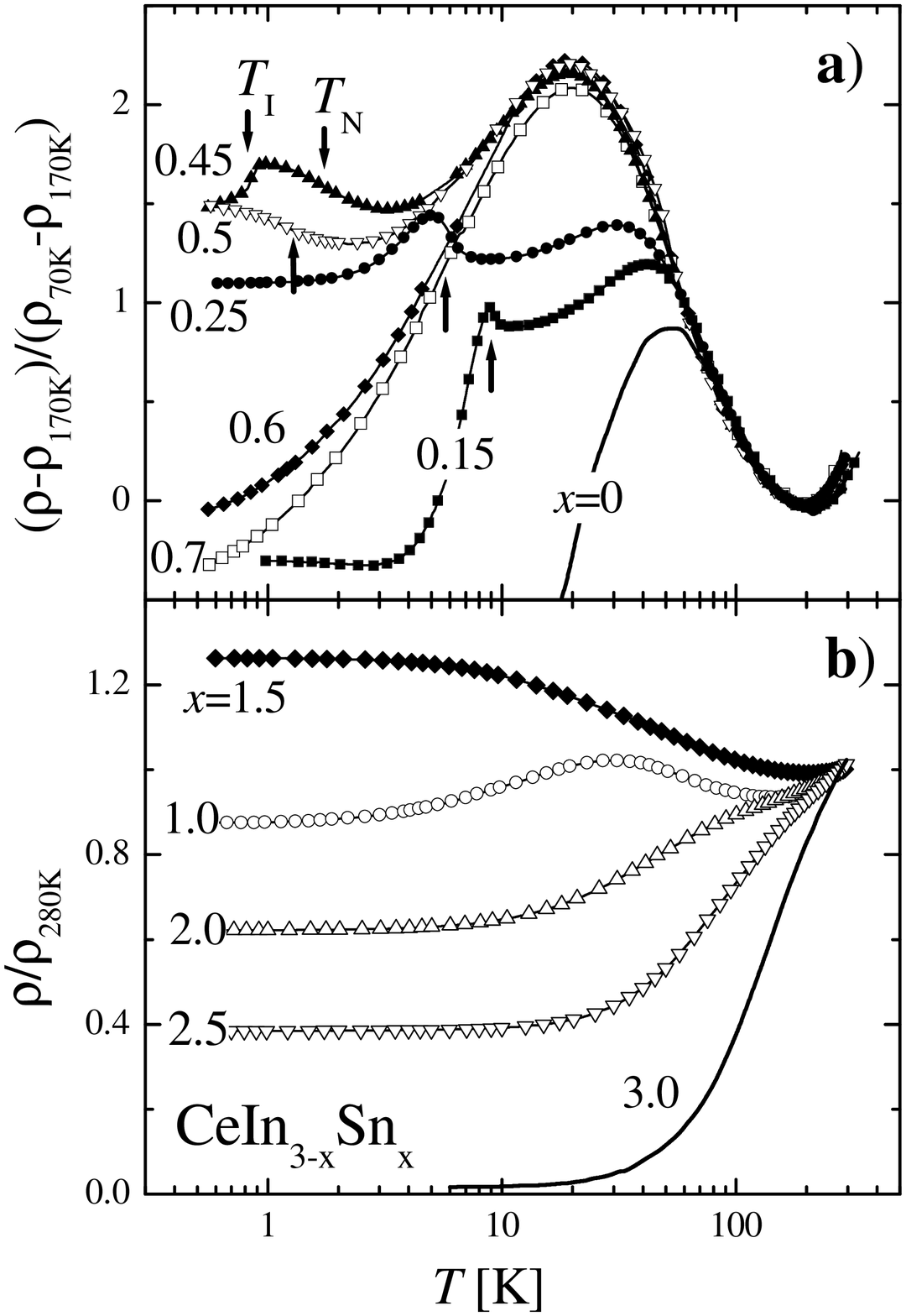}
\caption[]{Normalized electrical resitivity of CeIn$_{3-x}$Sn$_x$ in a logarithmic temperature representation: a) for the In-rich concentrations with a normalization at 70\,K and 170\,K (see the text) and b) for the $x\geq 1$ concentrations normalized at 280\,K. Data for CeIn$_3$ and CeSn$_3$ are taken from references \cite{Elenbaas80} and \cite{Sereni80}, respectively. In panel a) the arrows indicate the AF transition as determined from $\chi(T)$ and $C_{\rm P}(T)$ measurements. For $x=0.45$, the first order transition at $T_{\rm I}$ is also labeled.}
\end{center}
\label{fig:ro}
\end{figure}

The electronic contribution to the specific heat of the alloys included in the antiferromagnetic region (i.e. $x\le 0.7$) is depicted in Fig.~4 in a logarithmic $T$ scale after phonon subtraction.[10] The $x=0.15$ sample shows a jump in $C_{\rm el}(T)$ at the N\'eel temperature: $\Delta C_{\rm el}(T_{{\rm N}})\approx 12\,$J/mol K$^2$, as expected for a mean-field like transition. As $x$ increases, the transition first broadens (for $x=0.25$, 0.3 and 0.37) and then (for $x=0.41$, $0.45$, $0.47$, $0.5$ and $0.55$) $\Delta C_{\rm el}$ becomes sharper again, being much smaller and followed at $T>T_{\rm N}$ by a nearly logarithmic tail. At $x=0.6$ no transition is observed within our accessible temperature range. However, $C_{\rm el}(T)$ measurements performed down to lower temperature confirm the existence of a transition at $T_{\rm N}=0.40\,$K,\cite{Custers03} with similar characteristics to those observed in the $x=0.55$ sample. Above $T_{\rm N}$, the $x=0.6$ sample displays a logarithmic $C_{\rm el}(T)$ dependence within more than a decade of temperature, which can be described by the function $C_{\rm el}(T)/T=0.50\log(16/T)\,$J/mol\,K$^2$. A similar type of behaviour has been observed in other cerium systems, in particular in the doping-induced NFL CeCu$_{6-y}$Au$_y$ (see for example references \cite{Lohneysen99} and \cite{Sereni02}). The $x=0.7$ sample also shows a logarithmic $T$ dependence, although $C_{\rm el}(T)/T$ flattens at low temperature. Specific heat measurements presented in Ref.~\cite{Custers03} show that this curvature extends down to lower temperature with a $C_{\rm el}(T)/T\propto \gamma_0-b\sqrt{T}$ dependence (after the substraction of a hyperfine nuclear contribution due to In). This behaviour is expected for a three dimensional antiferromagnet.\cite{Moriya95} We also include in Fig.~4 the $C_{\rm el}(T)$ results for the $x=1$ sample, which show a drastic decrease of $C_{\rm el}/T(T\to 0)$. Using a Kondo-impurity model,\cite{Desgranges82} the $\gamma$ value extracted from this measurement corresponds to $T_{\rm K}=45\,$K.

\begin{figure}[htb]
\begin{center}
\includegraphics[angle=270,width=.47\textwidth]
{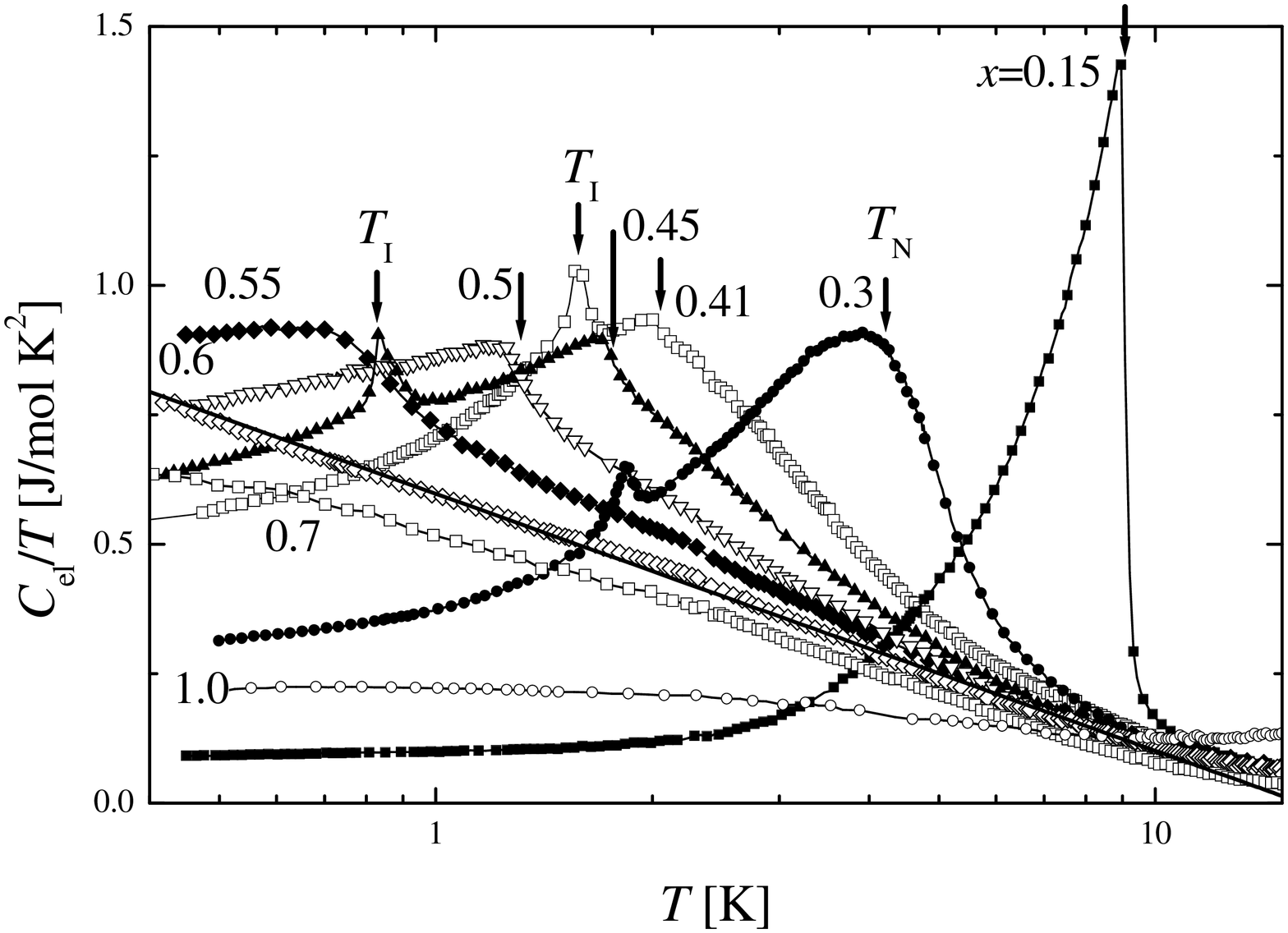}
\caption[]{Electronic specific heat of some representative samples with $x\le 1$ in a logarithmic temperature scale. The line is a fit to the function $C_{\rm el}(T)/T=D\log(T^*/T)$, see the text. The arrows show the position of $T_{\rm N}$. For $x=0.41$ and $x=0.45$, the first order transition at $T_{\rm I}$ is also labeled.}
\end{center}
\label{fig:cplog}
\end{figure}

The specific heat data of the samples with $x\geq 1$ are presented in Fig.~5, together with the data corresponding to the LaIn$_2$Sn sample which serves as phonon reference in this concentration range. The low temperature $C_{\rm P}/T$ values of these samples are strongly reduced when compared to those in the In-rich region. There is also a flattening of the curves at low temperatures, indicating the recovery of Fermi-liquid behaviour. However, a weak increase of $C_{\rm P}(T)/T$ is still present in the $x=1$ sample below 5\,K. The Sommerfeld coefficient in this region has a maximum value at $x=2$, where it reaches $\gamma=270\,$mJ/mol K$^2$, and then decreases monotonously down to 73\,mJ/mol K$^2$ for CeSn$_3$.\cite{Ikeda82} These results are in agreement with the $\gamma$ values reported in the literature,\cite{Elenbaas80} though no $C_{\rm P}(T)/T$ dependences were given there.

\begin{figure}[htb]
\begin{center}
\includegraphics[angle=270,width=.47\textwidth]
{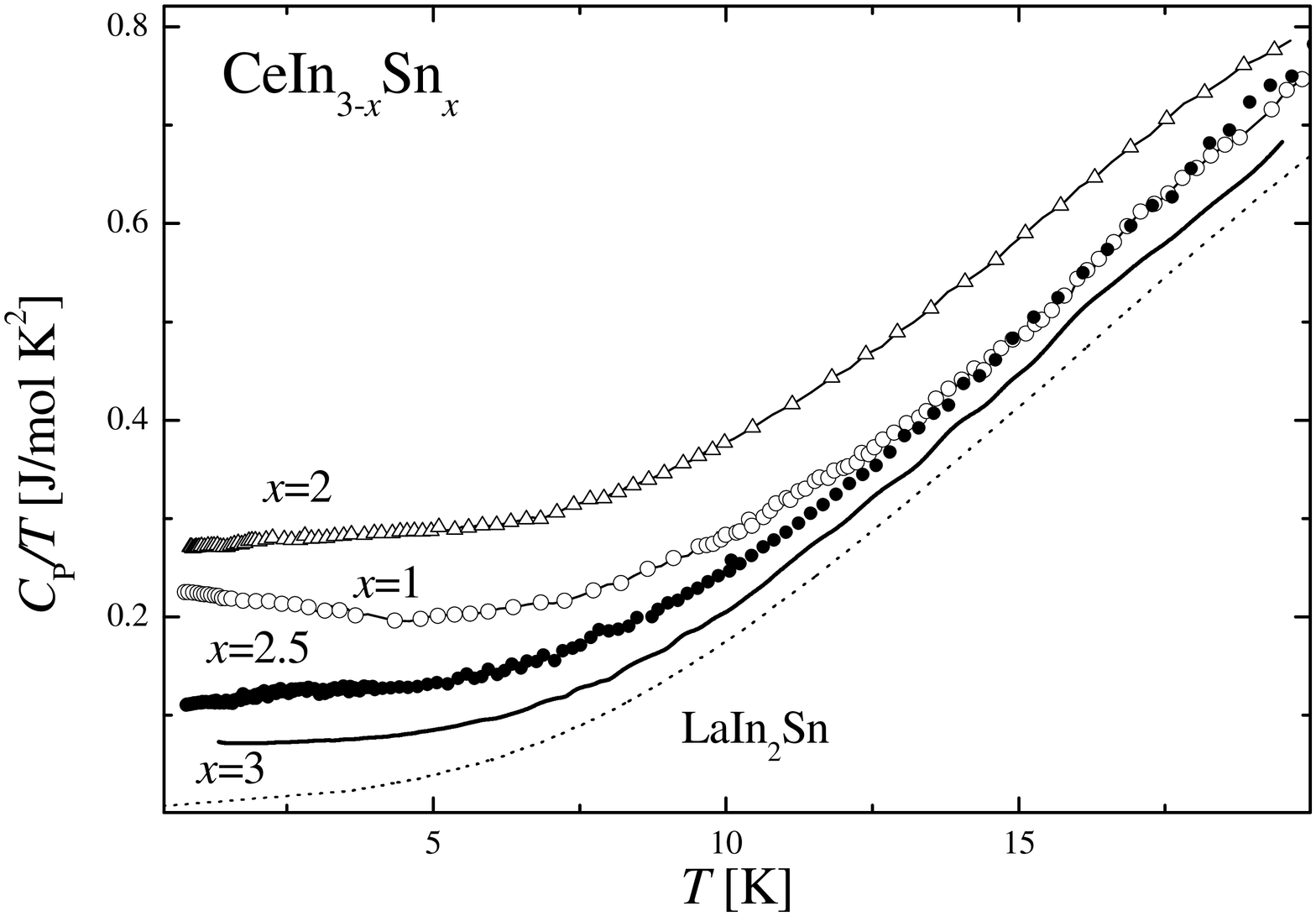}
\caption[]{Measured specific heat of samples with $x\ge 1$ in a $C_{\rm P}/T$ vs.~$T$ representation, including the results for CeSn$_3$ taken from Ref.~\cite{Ikeda82}. The data of LaIn$_2$Sn is included as a comparison (dotted line).}
\end{center}
\label{fig:cplinear}
\end{figure}

\section{Discussion}
\label{sec:discussion}

These experimental results indicate that the CeIn$_{3-x}$Sn$_x$ cubic-binary system is suitable for a detailed study of a magnetic to non-magnetic transformation which involves a quantum critical point. Because of the variety of properties to be considered, we will first discuss the high temperature results and low temperature ones will be discussed in the second part of this section.

Previous research performed by neutron scattering has shown that a precise estimation of the crystalline field splitting is difficult in this system because In is a strong neutron absorber which makes experiments rather difficult.\cite{Loewenhaupt} Taking profit of the simple CF level scheme of a cubic system, we can use our $\chi(T)$ data to estimate the crystal field splitting $\Delta_{\rm CF}$ between the $\Gamma_7$ doublet ground state (GS) and the excited $\Gamma_8$ quartet by fitting the measurements in the low-$x$ samples. In this fitting, we take into account the effects of a molecular field by introducing a $\lambda$-parameter and a temperature independent non-$4f$ contribution $\chi_0$:

\begin{equation}\label{eq:chicc}
\chi(T)=\frac{\chi_{\rm CF}(T)}{1+\lambda \chi_{\rm CF}(T)}+\chi_0.
\end{equation}

\noindent In this equation $\chi_{\rm CF}(T)$ corresponds to Van Vleck's formula, which in a cubic symmetry is a function of a single parameter, $\Delta_{\rm CF}$.\cite{Lueken79} By fitting the data of several samples above $10\,$K, we have calculated a mean value for the splitting: $\Delta_{\rm CF}=130\,$K, which coincides with the result of neutron scattering experiments performed on CeIn$_3$ ($\Delta_{\rm CF}\approx 12\,{\rm meV}=130\,$K \cite{Murani93}). 

The evolution of $\chi(T)$ with Sn concentration (see Fig.~2) can be separated into three regions. One close to $x=0$, where $\chi(T)$ is well described by equation \ref{eq:chicc}. Already at $x=0.7$ it is not possible to fit the $\chi(T)$ data with such a simple expression, probably due to the effect of the increasing Kondo temperature. In the intermediate region ($0.7<x<2$) the magnetic susceptibility $\chi_{2\,{\rm K}}$ drops down from its maximum at $x=0.45$ ($\chi_{2\,{\rm K}}\approx 17\times 10^{-3}\,$emu/mol) by  around 50\% due to a reduction of the moment of the Ce-doublet ground state ($\Gamma_7$). This crossover region ends with a weak increase of $\chi_{2\,{\rm K}}(x)$ at $x=2$. The concentration at which this relative maximum occurs coincides with the already quoted deviation of $V(x)$ from a linear behaviour (Vegard law), with the observation of a maximum in the $\gamma(x)$ dependence and with the beginning of a strong growth of $T^{\chi}_{\rm max}$. All these features have been related to a strong increase of hybridization effects as the system approaches the unstable valence state.\cite{Sereni95}

Concerning the resistivity measurements, the continuous reduction of $T^{\rho}_{\rm max}$ between CeIn$_3$ and $x\approx 0.37$, from $55\,$K down to $20\,$K, indicates a decrease of the temperature at which the coherence sets in. This strong decrease cannot be attributed to a reduction of the $4f$-conduction state hybridization because the Kondo temperature $T_{\rm K}$ (extracted from the entropy evolution, see below) increases, nor to a change in the crystal field splitting which would have to decrease by a $2.5$ factor between $x=0$ and $0.37$. This reduction of $T^{\rho}_{\rm max}(x)$ close to the stoichiometric compound could be produced by the disorder introduced in the alloying process, though it becomes concentration independent in the range $0.37\leq x \leq 0.7$. At higher Sn concentration, $T^{\rho}_{max}$ increases indicating a steeper increase of $T_{\rm K}$.

From the low temperature resistivity measurements, one observes that the $\Delta\rho=AT^2$ dependence is not detected above the critical concentration ($x_{\rm c}\approx 0.7$) within our experimental temperature range. This sub-quadratic power law behaviour of the resistivity is broken at $x=1.5$, with a low temperature $\Delta \rho(T)= \rho_0[1-(T/T_{\rm K})^2]$ dependence which indicates an impurity (instead of a lattice) behaviour of the magnetic moments. The $T_{\rm K}\approx 50\,$K value extracted from this formula is comparable to the one extracted for the $x=1$ sample (see the discussion of the entropy results in the following paragraph). However, in the high temperature region, the expected $\rho(T)\propto -\log(T)$ is not strictly observed, probably due to the crystal field excited level and phonon contributions. Although the low temperature coherent regime is recovered for $x>1.5$, it is only at $x=3$ where the $T^2$ dependence is clearly observed at least up to $40\,$K with $A=1.6\times 10^{-3} \mu\Omega\,{\rm cm}/{\rm K}^2$. The $n<2$ exponent observed in the resistivity for $x>x_{\rm c}$ may be due to the doping-induced disorder. However, we stress that the $n\approx 1$ exponent observed for $x\approx x_{\rm c}$ should be associated to an intrinsic feature around the antiferromagnetic QCP, since the $\Delta\rho\propto T^2$ temperature dependence is recovered under applied magnetic field.\cite{Custers03} 

The evolution of the entropy with increasing concentration provides valuable information on the variation of the low energy spectrum of excitations along the phase diagram. The entropy gained by the electronic system ($S_{\rm el}(T)$) is plotted in Fig.~6 for the samples with $x\le 1$. At low Sn concentration the transition at $T_{\rm N}$ results in a change of slope of $S_{\rm el}(T)$, as expected for a second-order phase transition. For $x=0.15$, the entropy above $T_{\rm N}=9.1\,$K is $S_{\rm el}(T_{\rm N},x=0.15)=0.8\,{\rm R}\ln(2)$. This value can be compared with $S_{\rm el}(T_{\rm N},x=0)\approx 0.87\,{\rm R}\ln(2)$, given for CeIn$_3$ in Ref.~\cite{Elenbaas80}. These results confirm that the doublet $\Gamma_7$ is involved in the magnetic order, lying well below the $\Gamma_8$ quartet. The deviation of the calculated entropy when compared to the expected value (${\rm R} \ln 2$) is probably due to a weak Kondo effect. As $x$ increases, less entropy is collected at some fixed temperature, meaning that the characteristic Kondo temperature is increasing. To estimate the evolution of $T_{\rm K}(x)$ we have calculated $\Delta S_{\rm el}^{12\,{\rm K}}\equiv S_{\rm el}(T=12\,{\rm K})$, plotted in the inset of Fig.~6. Between $x=0.15$ and $x=1$ this quantity diminishes continuously from $0.87\,{\rm R}\ln(2)$ down to $0.33\,{\rm R}\ln(2)$. By means of a Kondo impurity model \cite{Desgranges82} $T_{\rm K}(x)$ can be computed. We found that $T_{\rm K}(x=0.15)\approx 3\,$K and $T_{\rm K}(x=1)\approx 45\,$K. The value $\Delta S_{\rm el}^{12\,{\rm K}}$ for $x=1.5$ (also included in the inset of Fig.~6), corresponds to $T_{\rm K}(x=1.5)\approx 40\,$K. This value almost coincides with the Kondo temperature extracted from the fitting of the low temperature resistivity data. Such a constant behaviour of $T_{\rm K}$ between $x=1$ and $x=1.5$ is not expected within the Doniach diagram, where $T_{\rm K}$ rises exponentially with increasing hybridization. At that concentration, the eventual contribution of the excited quartet $\Gamma_8$ may explain this ``anomalous'' evolution of $T_{\rm K}$.

\begin{figure}[htb]
\begin{center}
\includegraphics[angle=270,width=.47\textwidth]
{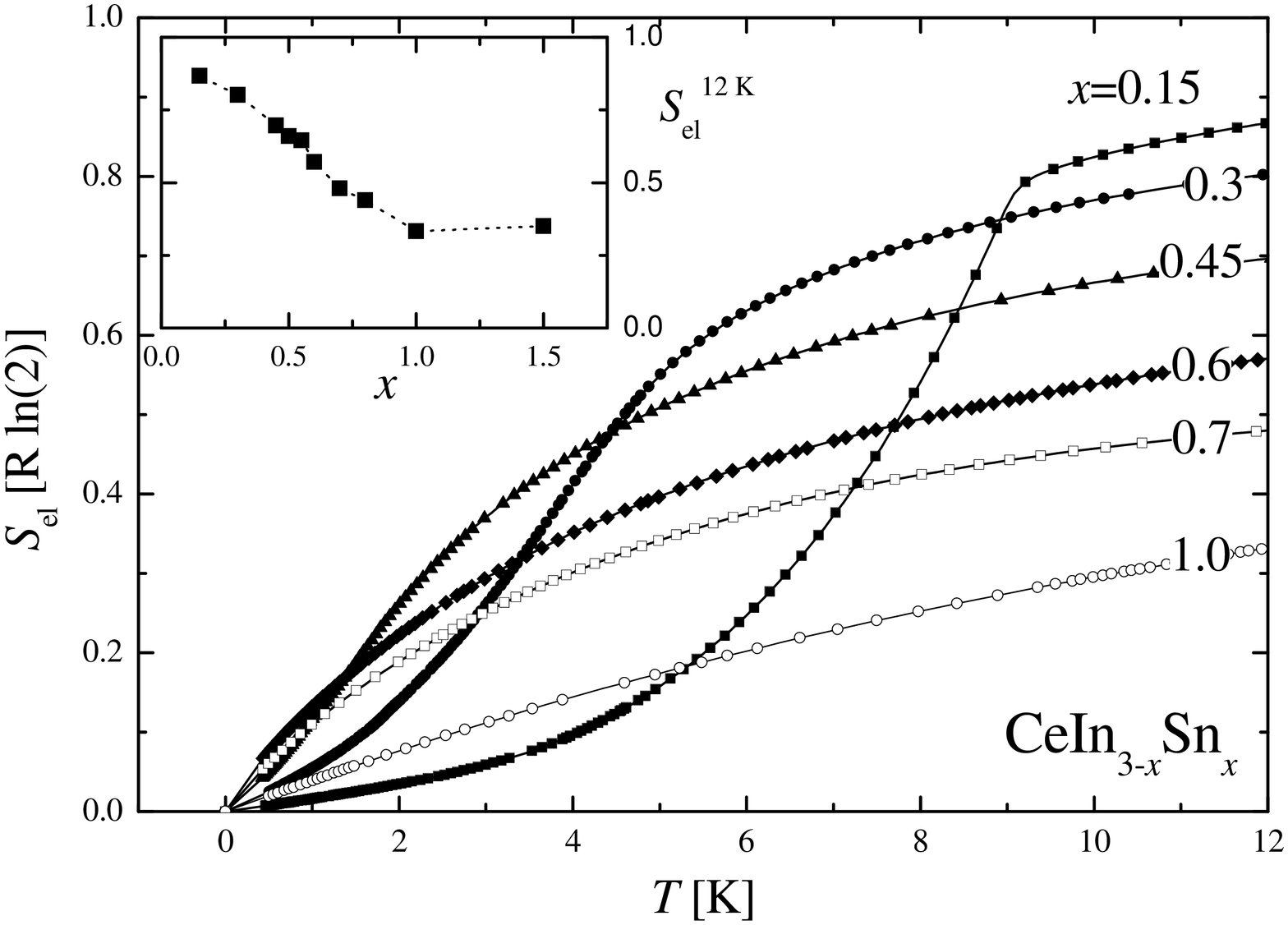}
\caption[]{Normalized electronic entropy $S_{\rm el}$ of some of the $x\leq 1$ samples. Inset: evolution of $\Delta S_{\rm el}^{12\,{\rm K}}(x)$ between $x=0.15$ and $1.5$ in ${\rm R}\ln(2)$ units. This value is employed to calculate the single impurity Kondo temperature (see the text).}
\end{center}
\label{fig:entropia}
\end{figure}

\begin{figure}[htb]
\begin{center}
\includegraphics[angle=270,width=.47\textwidth]{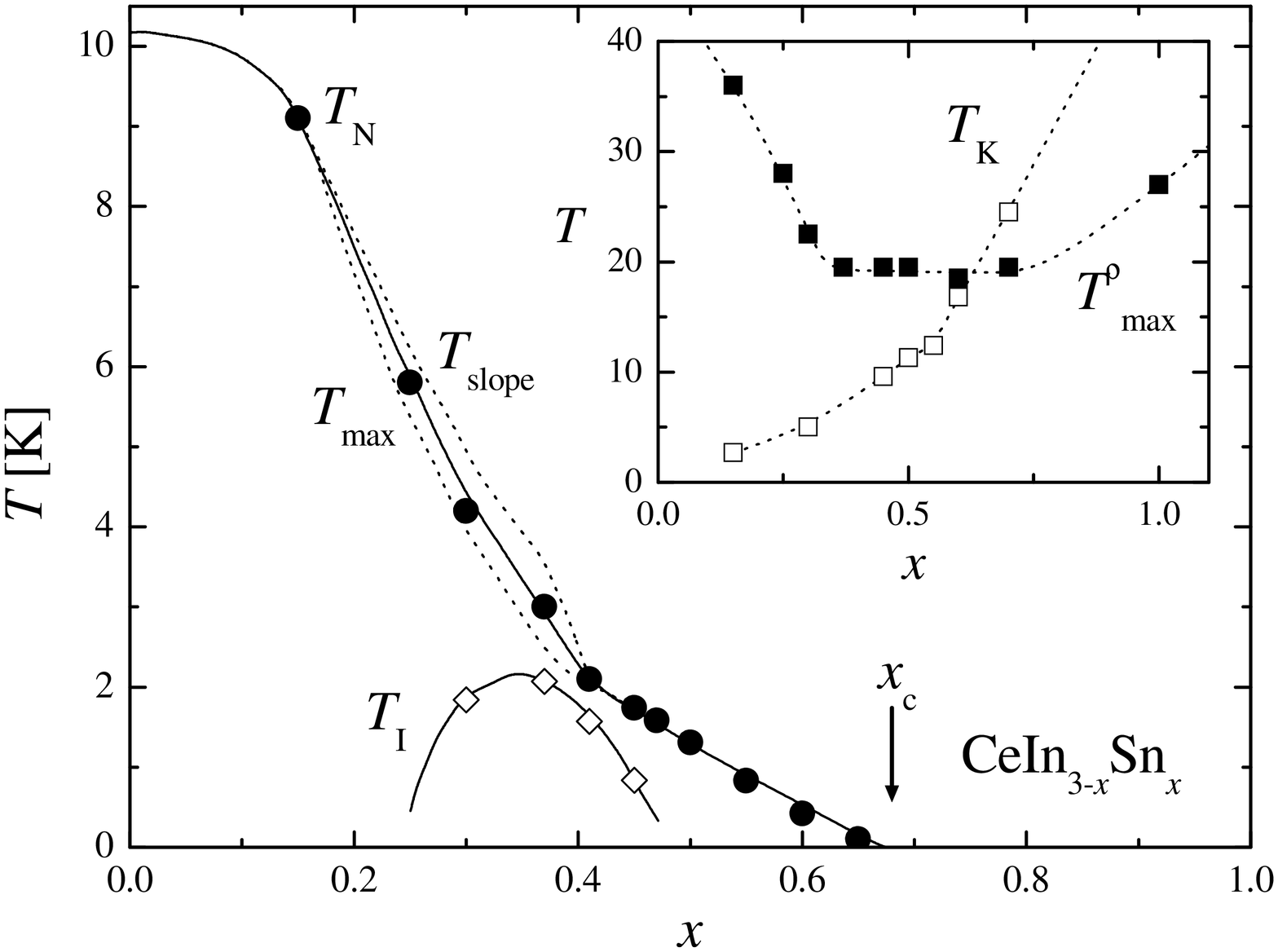}
\caption[]{Detailed magnetic phase diagram of CeIn$_{3-x}$Sn$_x$ for the $x\leq 1$ region (the $T_{\rm N}$ data of the $x=0.6$ and $x=0.65$ samples is taken from Ref.~\cite{Custers03}). The dotted lines represent the evolution of $T_{\rm max}(x)$ and $T_{\rm slope}(x)$, see the text. Inset: evolution of the Kondo temperature $T_{\rm K}(x)$ (as determined from a single impurity model) and the position of the maximum of the resistivity $T^{\rho}_{\rm max}(x)$.}
\end{center}
\label{fig:phasediag}
\end{figure}

The $T_{\rm N}(x)$ dependence we observe indicates that the antiferromagnetic order disappears at a quantum critical point near $x \approx 0.7$. However, instead of a monotonous trend with negative curvature, our results indicate a complex behaviour around $x \approx 0.3$ (see Fig.~7). This change of regime is reflected in the specific heat as a deviation of $\Delta C_{\rm el}(T_{\rm N})$ from a mean field type behaviour and the formation of a long $C_{\rm el}/T$ tail at $T>T_{\rm N}$. The broadening of the $C_{\rm el}(T)$ anomaly around $T_{\rm N}$ can be characterized by comparing the temperatures $T_{\rm max}(x)\le T_{\rm N}(x)$, at which $C_{\rm el}/T$ has its maximum, and $T_{\rm slope}(x)\ge T_{\rm N}(x)$ where the largest (negative) slope of $C_{\rm el}/T$ is detected, see Fig.~7. Above $x=0.41$, $T_{\rm N}(x)$, $T_{\rm max}(x)$ and $T_{\rm slope}(x)$ merge again, decreasing linearly down to $T_{\rm N}=0.1$ at $x=0.65$. The new information obtained from the present study, together with the low temperature data from reference \cite{Custers03}, relocate the critical concentration around $x_{\rm c}\approx 0.7$ instead of $x=0.4$, extrapolated by J.~Lawrence from higher temperatures.\cite{Lawrence79}

In coincidence with the change of regime near $x=0.3$, a second transition (in this case, with first order character) develops at $T_{\rm I}<T_{\rm N}$ within the $0.30\le x\le 0.45$ range. The temperature $T_{\rm I}$ evolves in a non-monotonous manner: it increases up to 2.1\,K for $x=0.37$ (almost merging the $T_{\rm max}(x)$ curve) and then it decreases down to $T_{\rm I}(x=0.45)\approx0.8\,$K at $x=0.45$. No signs of $T_{\rm I}$ where found in the $x=0.47$ sample within our temperature range. The merging of both transition lines around $x=0.4$ raises the possibility of a tetracritical point around this concentration as discussed in Ref.~\cite{Pedrazzini03}.

The change of the $T_{\rm N}(x)$ dependence at $x\approx 0.4$ could be related to the onset of NFL behaviour, clearly before reaching the critical concentration. Such a behaviour is reflected in the almost logarithmic tail of the specific heat, in the drastic change of the $C_{\rm el}/T$ anomaly associated to the AF order and in the onset of a linear dependence of $T_{\rm N}(x)$. Whether the simultaneous observation of a first order transition is related to this fact cannot be established with the available experimental data.

The effect of Sn doping on CeIn$_3$ is to be contrasted with that produced by hydrostatic pressure, $p$. Although in both cases the weakening of the $4f$ moments with increasing control parameter ($p$ or $x$) is observed, the unit-cell volume $V(p)$ and $V(x)$ have opposite slopes. This fact implies that Sn doping substantially modifies the electronic structure of CeIn$_3$ as the electron-count changes. As a consequence, both phase diagrams show striking differences besides a continuous reduction of $T_{\rm N}$: no evidence of superconductivity appears in the alloy despite the low temperature to which $T_{\rm N}$ was detected. On the other hand, the first order transition at $T=T_{\rm I}$ has not been observed in CeIn$_3$ under pressure. Moreover, the features observed in $\rho(T)$ around $T_{\rm N}$ or the exponent of the $\rho(T)\propto T^n$ dependence close to the critical region are quite different when controlling the two different parameters.\cite{Knebel01}

\section{Conclusions}

This study shows that the low temperature magnetic phase diagram of CeIn$_{3-x}$Sn$_x$ is much richer than expected from previous works. Although the ordering temperature decreases continuously with increasing Sn content towards $T_{\rm N}=0\,$K at $x_{\rm c}\approx 0.7$, the evolution of the antiferromagnetic state is not monotonous. Instead, our results indicate that around $x\approx 0.4$ there are clear changes in the concentration dependencies of the ordering temperature, specific heat anomaly and electrical resistivity. Those changes also affect the paramagnetic phase, where the logarithmic behaviour $C_{\rm el}/T \propto -\log(T)$ at $T>T_{\rm N}(x)$ extends even above the N\'eel temperature of stoichiometric CeIn$_3$, $T_{\rm N}(x=0)=10\,$K. For $x>0.4$, $T_{\rm N}(x)$ decreases linearly down to $T_{\rm N}=0.4\,$K at $x=0.6$. The decrease of $T_{\rm N}(x)$ over more than an order of magnitude in temperature (i.e., from 10\,K down to 0.4\,K), exceeds the temperature range of $T_{\rm N}(p)$ observed in stoichiometric CeIn$_3$ under hydrostatic pressure.\cite{Knebel01,Kawasaki02} To our knowledge, the decrease of $T_{\rm N}$ in such an extended temperature range has not previously been reported. This fact indicates that in CeIn$_{3-x}$Sn$_x$ the disorder introduced by doping is not the dominant factor in the evolution of the characteristic parameters such as the magnetic exchange or the Kondo screening. As a consequence, the CeIn$_{3-x}$Sn$_x$ phase diagram up to $x=1$ can be described within the Doniach scheme.

Due to the cubic point symmetry at the Ce site, this research proves that structural anisotropy is not determinant for the appearance of NFL behaviour. In spite of this, the $T$ dependences observed in this research (almost linear $\rho(T)$ and logarithmic $C_{\rm el}/T$ for $T>0.4\,$K) would imply a low dimensional scenario for this NFL. Eventhough the low temperature measurements suggest the expected $C_{\rm el}/T\propto\sqrt T$ at lower temperatures, the $\rho(T)\propto T$ observed in Ref.~\cite{Custers03} differs with the temperature dependence expected by current models for a three dimensional antiferromagnet.

Around $x=2$, the anomalous concentration dependence of the lattice parameter, of $\chi_{2\,{\rm K}}$ and $C_{\rm el}/T$ (at $T=0.5\,$K) can be explained by a non-monotonous evolution of the hybridization strength with concentration as suggested in Ref.~\cite{Lawrence79}, and by the evolution of the system towards the formation of the IV state.\cite{Sereni95} In this limit the ground state degeneration is expected to involve the six fold $J=5/2$ level. Further progress on the knowledge of the nature of the AF phase close to the critical concentration and the phase formed below $T_{\rm I}$ requires  microscopic measurements, which are presently under way. 

\begin{acknowledgement}

This work was supported by a DAAD (Germany) and F.\ Antorchas (Argentina) cooperation program (Pr.\ Nr.\ 13740/1-88). The authors P.\ P., M.\ G.\ B.\ and J.\ G.\ S.\ are affiliated to CONICET.

\end{acknowledgement}



\end{document}